\documentclass[11pt]{article}  
\usepackage{ltexpprt} 
\usepackage{algorithm,algorithmic,amsmath,graphicx,amsfonts}
\usepackage{xspace}
\newtheorem{observation}[lemma]{Observation}

\DeclareMathOperator{\cost}{cost}
\DeclareMathOperator{\entropy}{{\cal H}}
\newcommand{\OPT}{\mbox{\sc opt}\xspace}

\newcommand{\opt}{\OPT}
\newcommand{\FCFS}{\mbox{\sc fcfs}\xspace}
\newcommand{\Fcfs}{\mbox{\sc Fcfs}\xspace}
\newcommand{\fcfs}{\FCFS}
\newcommand{\fcfsu}{\ensuremath{\mbox{\sc fcfs}_\calU}\xspace}

\newcommand{\Rp}{\mathbb{R}_{+}}

\newcommand{\calI}{{\cal I}}       
\newcommand{\calA}{{\cal A}}    
\newcommand{\calU}{{\cal U}}     
\newcommand{\calS}{{\cal S}}     
\newcommand{\calL}{{\cal L}}     

\newcommand{\ifsketch}[2]{#2}

\newcommand{\qed}{\ensuremath{\Box}}
\newenvironment{Proof}{\begin{proof}}{\hfill\qed\end{proof}}

\title{First Come First Served\\ for Online Slot Allocation and Huffman Coding}
\author{
  Monik Khare
  \and
  Claire Mathieu\thanks{\'Ecole Normale Sup\'erieure, Paris, France.  Research partially supported by NSF grant CCF-0964037.}
  \and
  Neal E. Young\thanks{University of California, Riverside, California, U.S.A. Research partially supported by NSF grant CCF-1117954.}
}
{\date{\em full version with full proofs\footnote{Extended abstract appears in SODA 2014.}}}

\begin{document}

\maketitle

\begin{abstract}
Can one choose a good Huffman code on the fly, without knowing the underlying distribution?
{\em Online Slot Allocation (OSA)} models this and similar problems:
There are $n$ slots, each with a known cost.
There are $n$ items.
Requests for items are drawn i.i.d.~from a fixed but hidden probability distribution $p$.
After each request, if the item, $i$, was not previously requested,
then the algorithm (knowing $c$ and the requests so far, but not $p$)
must place the item in some vacant slot $j_i$, at cost $p_i \,c(j_i)$.
The goal is to minimize the total cost $\sum_{i=1}^n p_i\,c(j_i)$.

The optimal offline algorithm is trivial: 
put the most probable item in the cheapest slot,
the second most probable item in the second cheapest slot,
etc.
The optimal online algorithm is {\em First Come First Served} (\fcfs):
put the first requested item in the cheapest slot,
the second (distinct) requested item in the second cheapest slot, etc.
The optimal competitive ratios for any online algorithm are 
$1+H_{n-1} \sim \ln n$ for general costs and 2 for concave costs.  
For logarithmic costs, the ratio is, asymptotically, $1$: 
\fcfs gives cost $\opt + O(\log \opt)$.  

For Huffman coding, \fcfs yields an online algorithm (one that allocates codewords on demand, without knowing the underlying probability distribution)
that guarantees asymptotically optimal cost: at most $\opt + 2\log_2(1+ \opt)+2$.
\end{abstract}



\section{Introduction}\label{sec:intro}
Modeling an algorithm by its worst-case performance can be overly pessimistic.
Modeling by average-case performance 
on a specific input distribution can be overly optimistic.
A natural compromise is modeling by average-case performance 
on an adversarially chosen distribution:
a good algorithm should perform well on inputs drawn from
any distribution in some large class.
This ``worst-distribution'' approach is an emerging direction
in the study of online algorithms,
where standard (worst-case) competitive analysis 
can lead to overly pessimistic competitive ratios.

Here we introduce a online problem
that we call {\em Slot Allocation (OSA)}.
An instance is specified by $n$ {\em slots}, 
with (known) respective costs $c(1) \le c(2) \le \cdots \le c(n)$,
and a (hidden) probability distribution $p$ on $n$ items.
Requests for items are drawn i.i.d.~from $p$.
If the requested item has not been requested before,
the algorithm must assign the item $i$ to some vacant (not-yet-assigned) slot, $j_i$,
at cost $p_i \,c(j_i)$.
Play stops once all items have been requested.
The objective is to minimize the assignment cost, that is, $\sum_{i=1}^n p_i\, c(j_i)$. 
An {\em online} algorithm must choose each slot $j_i$
as a function of just the slot costs $c$ and the items chosen so far
(but not $p$).

The optimal offline solution is trivial: for each slot $j\in[n]$,
put the $j$th most probable item in slot $j$.
There is one natural online algorithm,
which we call {\em First Come First Served} (\fcfs):
put the $j$th (distinct) item requested in slot $j$.

The {\em cost} of a randomized algorithm on $(p,c)$
is its expected assignment cost (over all random choices of the algorithm
and draws from $p$).
The {\em competitive ratio} of an algorithm for a set $X$ of inputs,
is the supremum, over instances $(p,c)\in X$,
of the algorithm's cost on $(p,c)$ 
divided by the optimal offline cost for $(p,c)$.
We show that \fcfs has the minimum competitive ratio of any online algorithm.
We also determine the optimal ratio for various classes of costs:
it is
$1+H_k$ for general costs
(here $k \le n-1$ is the number of non-maximal coefficients in the cost vector)
and 2 for concave costs.
For logarithmic costs,
the asymptotic competitive ratio\footnote
{Formally, we define the asymptotic competitive ratio 
to be the limit, as $h\rightarrow\infty$,
of the competitive ratio against inputs 
where the entropy of $p$ is at least $h$.
For general costs and for concave costs
the optimal asymptotic ratios equal the non-asymptotic ratios.
This can be shown by adapting the lower-bound proofs
in \S\ref{sec:greedy lower bounds}.}
is 1:
\fcfs gives cost $\OPT+O(\log \OPT)$.

We apply \fcfs to \emph{Online Huffman Coding (OHC)}.
For OHC the offline problem is standard Huffman coding:
given probability distribution $p$,
find a prefix-free code $\chi$ over $\{ 0,1 \}$ minimizing
the average codeword length, $\sum_{i=1}^n p_i |\chi_i|$~~\cite{huffman1952method}.
In the online problem,
requests are drawn i.i.d.~from $p$ and revealed one by one.
In response to each request, if the item $i$ has not been requested before,
the algorithm (knowing only the requests so far, but not $p$) 
must commit to its codeword $\chi_i$ for $i$.
(If the requests were ordered adversarially,
rather than drawn from $p$, no online algorithm could have bounded competitive ratio.)


Applying \fcfs to OHC yields an algorithm that guarantees cost at most
$\opt + 2\log_2 (1+ \opt) + 2$.
The algorithm uses a so-called {\em universal codeword set~\cite{elias1975universal,golin2012huffman}}\nocite{golin2002huffman}
to relax the prefix-free constraint.
This makes OHC a special case of OSA;
applying \fcfs gives the algorithm.

\subsection*{Related work.}
Analyzing \fcfs requires analyzing
sampling without replacement from a non-uniform distribution.
This poorly understood process (see e.g.~\cite{fog2008calculation,fog2008sampling})
underlies the independent chip model
(for valuing chips in poker tournaments, e.g.~\cite{gilbert2009independent})
and Wallenius' noncentral hypergeometric distribution
\cite{wallenius1963biased}.

\paragraph{Adaptive Huffman coding.}
The so-called {\em adaptive Huffman coding} algorithm
also requires no a-priori knowledge of the frequency distribution,
but the algorithm modifies the code online:
the $i$th item in the sequence is transmitted using
the codeword from the optimal Huffman code for just the first $i-1$
items (except for first occurrences)~\cite
{faller1973adaptive,gallager1978variations,knuth1985dynamic,vitter1985design,vitter1987design,vitter1989dynamic}.
Adaptive Huffman coding gives cost at most $\OPT+1$.
Both online and adaptive Huffman coding require only one pass
(regular Huffman coding requires two)
and nearly minimize the cost,
but online Huffman coding does so using a fixed (non-adaptive) code.  

\paragraph{List management and paging.}
These two online problems
\nocite{sleator1984amortized}\cite{borodin1998online,Sleator:1985}
are similar to OSA with, respectively, linear and 0/1 costs.
The differences are that in OSA
(a) the requests are drawn i.i.d.~from a distribution,
and (b) the algorithm is compared to the static \OPT,
which fixes the list order or cached set at the start and then doesn't change it.
For paging, the OSA model is essentially the {\em independent reference model (IRM)},
which has a long history, e.g.~\cite{aho1971principles,franaszek1974some}
and is still in use.
For list management, the OSA model was used by Rivest~\cite{rivest1976self}
in a precursor to the worst-case model~\cite{Sleator:1985}\nocite{sleator1984amortized}.
For list management, \fcfs orders items by first request;
for paging, \fcfs fixes the first $k$ (or $k-1$) requested items in the cache.
Our Thm.~\ref{thm:greedy upper bounds}
implies that these simple static strategies are, respectively, 2-competitive 
and $(1+H_k)$-competitive against their static \opt{}s.)

Other more sophisticated worst-distribution models for paging have been studied
\nocite{karlin1992markov,koutsoupias1994beyond,young1998bounding}
\cite{karlin2000markov,koutsoupias2000beyond,raghavan1992statistical,young2000line}.

In {\bf online bin packing},
items with sizes arrive online to be packed into $n$ unit-capacity bins. 
If sizes are i.i.d.~from a discrete distribution,
an online algorithm achieves cost $\opt +O(w(n))$, where $w(n)$ 
is $\log n$ for distributions with {\em bounded waste} 
and $\sqrt n$ for those that are {\em perfectly packable}~\cite{CsirikJKOSW06,Gupta2012Online}.
In {\bf online knapsack}, 
items with weights and values arrive online;
each must be packed or rejected on arrival.  Lueker studied a model where
items are drawn randomly from any distribution in a large class~\cite{Lueker:1995}.
In {\bf online facility location},
points arrive online; the algorithm opens facilities to serve them. 
The worst-case ratio is $\Theta(\log n)$;
for randomly ordered points,
the competitive ratio is $O(1)$~\cite{Meyerson01}.
In {\bf online Steiner tree},
given a graph, terminal vertices arrive online
and must be connected on arrival by adding a path to the current tree.
The worst-case ratio is $\Theta(\log n)$,
but if terminals are i.i.d.~from a known distribution,
the ratio is $O(1)$~\cite{Meyerson01}.

In {\bf the secretary problem},
a random permutation of ranked items is revealed item by item.
The algorithm must reject or select each item as it is revealed.
Only one item can be selected; the goal is to choose the top-ranked item~\cite{Dynkin1963,Gilbert1966}.
Similarly to OSA, the offline problem is trivial but the (random) online setting is interesting.
Random online models for many variations of the secretary problem have been studied recently, mainly in the larger context of {\bf online auctions}~\cite{Babaioff:2008}.  (See~\cite{Ferguson1989,Freeman1983} for older work.)
In the {\bf adwords} problem, given a known set of keywords, advertisers arrive online.
Each advertiser comes with a subset of keywords,
the algorithm must allocate an unused one to the advertiser (if any are still available).
The objective is to maximize the number of assigned keywords.
If the advertisers arrive in random order,
the optimal competitive ratio is $0.696$ or more~\cite{Mahdian:2011,karande2011},
better than the worst-case ratio of $1-1/e \approx 0.632$~\cite{Karp1990}. 
Other random online models, including hidden distributions, have been studied too~\cite{Devanur:2011}.

\subsection*{Open problems.}  
For OSA with logarithmic costs and OHC,
what are the optimal non-asymptotic competitive ratios?
Are the $O(\log\opt)$ additive gaps in our upper bounds tight?
For problems such as paging, list management, and adaptive Huffman Coding 
(which allow dynamically adjusting the current online solution at some cost),
for various ``worst-distribution'' models,
what are the worst-case gaps between the costs of the static \opt and the dynamic \opt?

\subsection*{Preliminary.} 
In the body of the paper, for convenience,
we relax the assumption that $p$ is a probability distribution
(i.e., that $\sum_{i=1}^n p_i = 1$), replacing $p$ by any
frequency distribution $f\in \Rp^n$.
(To sample from $f$, use the normalized distribution $f/\sum_{i=1}^n f_i$.)
Also, $f(i)$ and $c(j)$ are synonymous with $f_i$ and $c_j$, respectively.

\section{FCFS is optimal for Online Slot Allocation}\label{sec:greedy optimal}

In this section we prove that the First-Come-First-Serve algorithm (\fcfs) is optimally competitive for OSA:
\begin{theorem}\label{thm:greedy optimal}
\fcfs is optimally competitive
for any class $X$ of OSA inputs that is closed under permutation of the frequency distribution
(that is, for any $(f,c)\in X$, and any permutation $\pi$ of $[n]$,
the instance $(f',c)$ is also in $X$, where $f'(i) = f({\pi_i})$).
\end{theorem}

Intuitively, the theorem is not surprising, but the proof is technical
(due to the unbounded depth of the game tree) and indirect.
\ifsketch{Some parts of the proof are only sketched here.  
Missing details are in the full paper
(a preliminary version is available \cite{khare2013first}).}{}

For the proof we use the following formal definition of an algorithm for OSA.
Define an {\em algorithm state} $u=(f,c,(i_1,\ldots,i_t),g,i)$
to be a tuple where $(f,c)$ is the OSA instance, 
$(i_1,\ldots,i_t)$ are the first $t$ requests,
$g$ is an assignment of slots to requested items,
and $i$ is request $t+1$, which needs a slot
(so $i\not\in\{i_1,\ldots,i_t\}$).
Formally, a (randomized) algorithm $\calA$ is defined by,
for each algorithm state $u$,
a distribution $\calA(u)$ on vacant slots, 
from which the slot for item $i$ is chosen at random.

$\calA$ is online if, for all $u=(f,c,(i_1,\ldots,i_t),g,i)$,
distribution $\calA(u)$ is determined by
$(c,(i_1,\ldots,i_t),g,i)$ --- the state without $f$.

$\calA$ is {\em unbiased} if, for all $u=(f,c,(i_1,\ldots,i_t),g,i)$
distribution $\calA(u)$ is determined by $(f,c,(i_1,\ldots,i_t),g)$,
the state without $i$ (the ``name'' of the requested item).

$\calA$ is {\em stateless} if, for all $u=(f,c,(i_1,\ldots,i_t),g,i)$,
distribution $\calA(u)$ is determined by $(f,c,W,V)$,
where $W=[n]-\{i_1,\ldots,i_t\}$ is the set of items without slots
and $V=[n]-\{g(i_1),\ldots, g(i_t)\}$ is the set of vacant slots.

Any stateless algorithm $\calA$ is unbiased,
but $\calA$ can be stateless or unbiased without being online.
\Fcfs is online, unbiased, and stateless.

Here is a summary of the proof:
\begin{enumerate}
\item {\em For any online algorithm there is an equally competitive unbiased algorithm.}
This is simply by symmetry --- in the event that the requested item $i$ has not yet been requested, the ``name'' of the item gives no useful information,
because the frequency distributions are symmetric (closed under permutation of the item names).
\item {\em For any input $(f,c)$ and any unbiased algorithm,
some stateless algorithm achieves the same cost.}
An algorithm is stateless if it ignores repeat requests (requests to items that 
were previously requested).  Intuitively, repeat requests give no useful information,
because they do not change the part of the problem that remains to be solved.
\item {\em \Fcfs gives minimum cost for $(f,c)$ among stateless algorithms.}
Roughly, this is because, in expectation, the frequency of the first item chosen
is larger than that of the second (or any subsequent) item.
By a simple exchange argument, the optimal strategy may as well 
give the best slot (slot 1) to the first item chosen.
By induction, the optimal strategy may as well be \fcfs.
\end{enumerate}

\begin{lemma}\label{lemma:unbiased optimal}
For any online algorithm $\calA$, for any class of inputs 
that is closed under permutation of the frequency distribution,
there is an unbiased algorithm $\calA'$ 
that is equally competitive.
\end{lemma}
{
\begin{Proof}
  $\calA'$ randomly relabels the items
  then simulates $\calA$ on the relabeled instance.
  Relabeling makes $\calA'$ unbiased.  $\calA'$ is as competitive
  as $\calA$, because relabeling the items doesn't
  change \opt.  Here are the details.

$\calA'$, on a given instance $(f,c)$,
first renames the $n$ items according to a random permutation $\pi$,
then simulates $\calA$ item by item on the renamed instance $(f',c)$.
(Formally, on any algorithm node $u=(f,c,(i_1,\ldots,i_t),g,i)$,
the algorithm $\calA'$ uses the slot distribution 
$\calA'(u) = \calA(u')$,
where $u'=(f,c,(\pi(i_1),\ldots,\pi(i_t)),g_\pi,\pi(i))$
and $g_\pi$ is defined by $g_\pi(\pi(i)) = g(i)$.)

The random renaming of items ensures that $\calA'$ is unbiased. 
This is because, when an item $i$ is presented to $\calA'$ for the first time
(i.e., $i\in [n] - \{i_1,\ldots,i_{t}\}$),
the item $\pi(i)$ that $\calA'$ presents to $\calA$ is uniformly distributed
over the items not yet seen by $\calA$.
Therefore, $\calA'(u)$ is independent of
$u$'s particular item $i\in[n]-\{i_1,\ldots,i_{t}\}$.

$\calA'$ is as competitive as $\calA$, because,
for any instance $(f,c)$, 
the cost that $\calA'$ incurs
is at most the cost that $\calA$ incurs
on the permuted instance $(f',c)$, 
while the optimal cost for $(f',c)$ equals the optimal cost for $(f,c)$.
\end{Proof}
}
Recall that unbiased/stateless algorithms
``know'' $(f,c)$.  

\begin{lemma}\label{lemma:stateless optimal}
  Fix any instance $(f,c)$ of OSA.  For any unbiased algorithm $\calA$,
  there is a stateless algorithm $\calA'$ 
  whose cost on $(f,c)$ is at most the cost of $\calA$ on $(f,c)$.
\end{lemma}
\ifsketch{

The two-page proof of this lemma is in the full paper \cite{khare2013first}.   
The proof is technical due to the unbounded depth of the game tree,
but conceptually straightforward:
We fix $(f,c)$, then model the process of $\calA$ 
choosing slots for $i_1,i_2,i_3,\ldots$ (drawn i.i.d.~from $f$)
as an unbounded one-player game against chance, where each node of the game
tree represents a possible state of the process.
At any such node $u$, the (infinite) subtree $T_u$ rooted at that node represents the subgame
that remains to be played.
Since $(f,c)$ is fixed and known, but $\calA$ is unbiased,
the structure of subtree $T_u$ is determined (up to isomorphism) by the
pair $(W_u,V_u)$ --- the items without slots, and the vacant slots, in state $u$.
We say any two nodes that have the same pair $(W_u,V_u)$ are {\em equivalent},
and then argue that, for each of the (finitely many) equivalence classes,
$\calA$ may as well use the same strategy 
for each subtree $T_u$ rooted at any node in the equivalence class.
 Finally, the proof shows that if $\calA$ does so, then $\calA$ is stateless.
}
{
\begin{Proof}
  To reason precisely about running $\calA$ on $(f,c)$, we describe the process of choosing
  an allocation as a one-player game (against chance) in extensive form.
  To model that $\calA$ is unbiased, we change the process slightly:
  we make $\calA$ choose the next slot without knowing which unseen item
  the slot is for.  More specifically, we break the step of choosing the next item $i$
  into two steps.  First, we choose whether the next item will be taken from the seen items,
  or from the unseen items.  In the former case, we immediately 
  choose the next item from the seen items.  In the latter case, before choosing the next item,
  we consult the algorithm $\calA$ to determine the slot that the item will be assigned,
  and then choose the item after determining its slot.

  The game tree $T$ for the game that representing an instance $(f,c)$ has three types of nodes,
  each identified with a possible state of the process.
  Each non-leaf node has edges to its children,
  representing the possible transitions to the next state.
  \begin{itemize}
    \item A {\sf draw}  node represents the state just before the next item is drawn.
      The node is a tuple $u=((i_1,i_2,\ldots,i_t), g)$,
      where $i_1,\ldots,i_t$ is the sequence of items drawn so far,
      and $g:\{i_1,\ldots,i_t\}\rightarrow [n]$ 
      is the (injective) allocation slots to drawn items.
      The root node of $T$ is a draw node $u=((), g)$ where $g$ is the empty allocation.

      If all items have been seen (i.e., $\{i_1,\ldots,i_t\} = [n]$), then the draw node $u$ is a leaf.
      Otherwise, the edges out of the node are as follows.
      For each seen item $i$,
      there is an edge labeled $i$ to the draw node $w=((i_1,i_2,\ldots,i_t,i), g)$.
      This edge is followed with probability proportional to $f_i$.
      Following the edge corresponds to drawing item $i$ as the next item in the sequence.

      For the group of unseen items, there is a single edge to
      the {\sf choose} node $w=((i_1,i_2,\ldots,i_t), g, *)$.
      This edge is labeled $*$,
      and is followed with probability proportional 
      to the sum of the frequencies of the unseen items.
      Following this edge corresponds to committing to draw an unseen item
      as the next item (but not yet drawing the item).
  
    \item A {\sf choose} node represents the state just before the algorithm chooses
      the slot for the (yet to be chosen) unseen item.
      The node is a tuple $u=((i_1,i_2,\ldots,i_t), g, *)$,
      where $i_1,\ldots,i_t$ and $g$ are as for a draw node.
      For each unused slot $j$ (i.e., $j\in [n] - \{g(i_1),g(i_2),\ldots,g(i_t)\}$),
      there is an edge labeled $*\mapsto j$
      to the {\sf assign} node $w=((i_1,i_2,\ldots,i_t), g, j)$.
      The probabilities for the edges out of $u$ are unspecified
      (they will be determined by the strategy chosen to play the game;
      in other words, by the algorithm $\calA$).
      Following the edge corresponds to choosing the slot $j$ to be assigned
      to the next (to be chosen) unseen item.
      
    \item An {\sf assign} node represents the state just after the slot has been chosen,
      and just before the unseen item that will receive that slot has been chosen.
      The node is a tuple $u=((i_1,i_2,\ldots,i_t), g, j)$,
      where $i_1,\ldots,i_t$ and $g$ are as for a draw node, and $j\in[n]$ is a slot
      not used by $g$.
      For each unseen item $i$ (i.e., $i\in[n]-\{i_1,\ldots,i_t\}$),
      there is an edge labeled $i\mapsto j$
      to the draw node $w=((i_1,\ldots,i_t,i),g')$,
      where $g'$ is $g$ extended by $g'(i) = j$ (assigning slot $j$ to $i$).
      This edge is followed with probability proportional to $i$'s frequency $f_i$
      (normalized by the sum of the frequencies of the unseen items).
      Taking the edge corresponds to assigning slot $j$ to item $i$.
    \end{itemize}  
This defines the game tree $T$ for $(f,c)$.
Formally, the game is a one-player game against chance,
where the {\sf draw} and {\sf assign} nodes are chance nodes,
and the {\sf choose} nodes belong to the player.
The payout (at each leaf node) is the cost of the final allocation $g$.

A (behaviorally randomized) {\em strategy} $A$ for the player
assigns, to every {\sf choose} node $u$, a distribution $A(u)$ on the
edges out of $u$.
Such strategies correspond bijectively to the unbiased algorithms $\calA$ for $(f,c)$. 
The strategy $A$ corresponding to $\calA$ can be determined as follows.
For {\sf choose} node $u=((i_1,\ldots,i_t),g,*)$,
choose any unseen item $i$,
let $u'$ be the algorithm state $(f,c,(i_1,\ldots,i_t),g,i)$,
and take $A(u) = \calA(u')$, the slot-distribution for $u'$.
Since $\calA$ is unbiased, this distribution is independent of the choice of $i$.

Say that a strategy $A$ is {\em stateless} if
the distribution $A(u)$ at each {\sf choose} node $u=((i_1,\ldots,i_t),g,*)$ 
depends only on the set $W=[n]-\{i_1,\ldots,i_t\}$ of unseen items 
and the set $V=[n]-\{g(i_1),\ldots,g(i_t)\}$ of vacant slots.
Call the pair $(W,V)$ the {\em configuration} at node $u$.
Crucially, the subtree $T_u$ of $T$ rooted at $u$
(including the leaf costs, all edge labels, and all probabilities
except for the probabilities assigned to the choose nodes by $\calA$)
is determined by the configuration.  Intuitively, this means that
the distribution $A(u)$ (and $A$'s action throughout $T_u$)
may as well depend only on the configuration.  
We show that this is true.
This will complete the proof,
if $A$ is stateless, it's corresponding algorithm $\calA$ is also stateless.

Fix any strategy $A$.  Modify $A$ as follows.  
Consider, one at a time, each equivalence class $X$ of {\sf choose} nodes,
where two {\sf choose} nodes are equivalent if they have the same configuration.
Note that if $u=((i_1,\ldots,i_t),g,*)$ and $u'$ are both in $X$,
then $T_u$ and $T_{u'}$ are disjoint
(because leaving node $u$ increases the set of seen items).
Consider the random experiment of playing the game once
(using $A$ to determine probabilities out of {\sf choose} nodes).
Letting random variable $x$ be the first node in $X$ (if any)
encountered, and abusing notation,
we can express the expected payout as
\begin{eqnarray*}
&&\Pr[x\mbox{ undefined}]\, E\big[\cost(A)\,|\, x\mbox{ undefined}\big]
\\ &&~+~
\sum_{u\in X} \Pr[x = u] \big[\cost(g_u) + \cost(A_u)\big]
\end{eqnarray*}
where $\cost(g_u)$ is the cost of the partial allocation $g$ at node $u$
and $\cost(A_u)$ is the expected cost of the remaining allocation
following node $u$.

Although $X$ is infinite,
by an averaging argument,
there must be a node $W\in X$ such that
\[\cost(A_W)~\le~
\frac
{\sum_{u\in X} \Pr[x =u] \cost(A_u)}
{\sum_{u\in X} \Pr[x =u]}.
\]
In other words,
\(\sum_{u\in X} \Pr[x =u] \cost(A_W)\) is at most \(\sum_{u\in X} \Pr[x =u] \cost(A_u)\).
Fix such a $W$.
For each node $u$ other than $W$ in the equivalence class $X$,
modify $A$ by replacing $A_u$ by $A_W$.
(Here $A_u$ represents $A$ restricted to the subtree $T_u$.
Recall that $T_u$ and $T_W$ are isomorphic.)
The modified strategy $A'$ has expected payout 
\begin{eqnarray*}
&&\Pr[x\mbox{ undefined}]\, E\big[\cost(A)\,|\, x\mbox{ undefined}\big]
\\ &&~+~
\sum_{u\in X} \Pr[x = u] \big[\cost(g_u) + \cost(A_W)\big].
\end{eqnarray*}
By the choice of $W$, this is at most the expected payout for strategy $A$.

After $A$ is modified in this way for every equivalence class of {\sf choose} nodes,
the resulting algorithm $A'$ is stateless.
(Note that the modification for one equivalence class
leaves $A'$ stateless within the previously modified equivalence classes.)
\end{Proof}
}

\smallskip
The next and final step is to show that, for any instance $(f,c)$,
\fcfs is optimal among stateless strategies.
For stateless strategies, the underlying one-player game against chance
has the following finite form:
{\em Generate a random permutation $i_1,i_2,\ldots,i_n$ of the items
by sampling from $f$ without replacement.
For each time $t=1,2,\ldots,n$, just before the next item $i_t$ is drawn,
choose the slot $j_t$ that will hold that item $i_t$.
Finally, pay the cost, $\sum_{t=1}^n f(i_t) c(j_t)$.}

Call this the {\em compact game} for $(f,c)$.

\begin{observation}\label{observation:compact}
  (a) The stateless algorithms for $(f,c)$ correspond to the strategies for the compact game.
  \\[1ex]
(b) The compact game has an optimal strategy that is deterministic
(i.e., each choice $j_t$ is completely determined by $i_1,\ldots,i_{t-1}$).
\end{observation}
Observation~\ref{observation:compact} (b) holds 
by a standard leaves-to-root induction
(because the game is a finite one-player game against chance).

The final step requires an additional observation.
In a permutation $i_1,\ldots,i_n$,
generated by drawing items without replacement from $f$,
items with larger frequency tend to be drawn earlier,
so, not surprisingly, given two adjacent indices $j$ and $j+1$,
the expected frequency $E[f(i_j)]$ of the $j$th item drawn
is as large as the expected frequency $E[f(i_{j+1})]$ of the $(j+1)$st item.
Indeed, this holds even if we condition
on the outcomes of all the other $n-2$ draws:

\begin{observation}\label{obs:decreasing expectations}
  Fix a frequency distribution $f(1) \ge \cdots \ge f(n)$.
  Let $i_1,i_2,\ldots,i_n$ be a random permutation of $[n]$ generated
  by sampling without replacement from $f$.
  Then, for any index $j\in[n-1]$,
  given any outcome $J$ of the $n-2$ random draws
  $i_1,\ldots,i_{j-1}$ and $i_{j+2},\ldots,i_n$ (all the draws except $i_{j}$ and $i_{j+1}$)
  the expectation of $f(i_{j})$ is at least that of $f(i_{j+1})$:
  \[ E[f(i_j)\,|\,J] ~\ge~ E[f(i_{j+1})\,|\,J].\]
\end{observation}
{
\begin{Proof}
  Fix any $J$.  We show \( E[f(i_j) - f(i_{j+1})\,|\,J] \ge 0\).

  Let $a$ and $b$ be the two slots not occurring in $i_1,\ldots,i_{j-1}$ or $i_{j+2},\ldots,i_n$,
  so there are two possible permutations consistent with $J$:
  one in which $(i_{j},i_{j+1}) = (a,b)$,
  the other in which $(i_{j},i_{j+1}) = (b,a)$.
  Call these two permutations $A$ and $B$, respectively, so that
  $E[f(i_j) - f(i_{j+1})\,|\,J]$, the expectation in question,  is
  \begin{eqnarray*}
    &&\Pr[A\,|\,J] \, \big( f(a) - f(b) \big) ~+~
  \Pr[B\,|\,J] \,\big( f(b) - f(a) \big)
  \\ && ~=~
  \frac{\Pr[A] - \Pr[B]}{\Pr[J]}\, \big( f(a) - f(b) \big).
\end{eqnarray*}
  Assume without loss of generality that $a<b$,
  so $f(a) \ge f(b)$.  To show the expectation
  is non-negative, we verify $\Pr[A] \ge \Pr[B]$.
  
  By calculation, the (unconditioned) probability that a given permutation
  $I = (i_1,\ldots,i_n)$ occurs is $\Pr[I] = \prod_{t=1}^n f(i_t)/S_{I}(t)$,
  where $S_{I}(t)$ is the tail sum $\sum_{s=t}^n f(i_s)$.

  Applying this identity to $A$ and to $B$, then canceling common terms,
  $\Pr[A]/\Pr[B]$
  is $S_B({j+1}) / S_A(j+1)$,
  which equals $[f(a) + S_B({j+2})] / [f(b) + S_A({j+2})]$.
  Since $f(a) \ge f(b)$ and $S_A({j+2}) =S_B({j+2}) \ge 0$,
  the ratio $\Pr[A]/\Pr[B]$ is at least 1.
\end{Proof}
}

Next we complete the final step:
\begin{lemma}\label{lemma:greedy optimal}
  Fix any instance $(f,c)$ of OSA.
  Among stateless algorithms,
  \fcfs gives minimum cost for $(f,c)$.
\end{lemma}
\begin{Proof}
  We show that \fcfs gives minimum cost
  among deterministic strategies for the compact game
  (this proves the lemma by Observation~\ref{observation:compact}).
  If $n\le 1$ the lemma is trivial.
  Assume $n \ge 2$ and fix any optimal deterministic strategy $A$ for the compact game.
  Let $j_1$ be the first slot that $A$ chooses for the instance $(f,c)$.
  Recall that $c_1 \le c_2 \le \cdots \le c_n$.

  After $A$ chooses its first slot $j_1$ (deterministically) and assigns
  it to the item $i_1$ drawn subsequently from $f$,
  the rest of the game corresponds to a compact game for an instance $(f',c')$ 
  of cardinality $n-1$, where $f'$ is obtained from $f$ by deleting item $i_1$,
  and $c'$ is obtained from $c$ by deleting slot $j_1$.
  By induction, the \fcfs strategy is optimal for that smaller instance.
  So (by modifying $A$, if necessary, but without increasing its cost) 
  assume that $A$ chooses all remaining slots in order of increasing cost
  (breaking ties among slots of equal cost by preferring slots with minimum index $j$).

  If slot $j_1$ is the minimum-cost slot $1$, then $A$ is \fcfs, and we are done.
  Otherwise, since $A$ plays \fcfs after the first item (and ties are broken consistently)
  $A$ chooses the minimum-cost slot among $[n]-\{j_1\}$ for the second item.
  This is slot 1.  That is, $A$ plays first slot $j_1$, and then slot 1.
  Consider the strategy $A'$ differs from $A$ only in that for the first two items
  $A'$ plays slot 1 and then slot $j_1$.
  The cost of $A'$ minus the cost of $A$ is
  \begin{eqnarray*}
    &&E\big[\,c(1) \, f(i_1) ~+~ c(j_1) \, f(i_2)\,\big]
    \\[2pt]
    &&~~ -  E\big[\,c(j_1) \,f(i_1)] ~-~ c(1) f(i_2)\,\big],
  \end{eqnarray*}
  which equals $\big(c(1) - c(j_1)\big) \, E[f(i_1) - f(i_2)]$.

  This is non-positive because $c(1) \le c(j_1)$
  and, by Observation~\ref{obs:decreasing expectations},
  $E[f(i_1)] \ge E[f(i_2)]$.
  Therefore, this algorithm $A'$
  also gives minimum expected cost.

  $A'$ plays slot 1 first.
  By induction on $n$, replacing $A'$s subsequent $n-1$ choices 
  by the choices that would be made by \fcfs 
  does not increase the expected cost.
  The resulting algorithm is \fcfs.
  Thus, \fcfs has minimum expected cost for $(f,c)$.
\end{Proof}

Lemmas~\ref{lemma:unbiased optimal},
\ref{lemma:stateless optimal},
and \ref{lemma:greedy optimal}
imply Thm.~\ref{thm:greedy optimal}.

\section{Competitive ratio of FCFS for OSA}\label{sec:greedy upper bounds}
\begin{figure*}[t]
  \centering
  \framebox{\parbox{\textwidth}
  {\setlength{\parindent}{0in} \setlength{\parskip}{1pt}\small
    {\sf ~~input:} {\em frequency distribution $f$ over $[n]$, partition of $[n]$ into two sets $U$ and $\overline U = [n]-\overline U$.}
   \\ {\sf output:} {\em random permutation $i_1,\ldots,i_n$ of $[n]$ that is distributed as if sampled without replacement from $f$}\sf
   \smallskip

  1. Draw random permutation $\pi = \pi_1,\ldots,\pi_{|U|}$ of $U$ by sampling without replacement from $f$ restricted to $U$.

  2. Draw random permutation $\overline \pi = \overline \pi_1,\ldots,\overline \pi_{|\overline U|}$ of $\overline U$ by sampling without replacement from $f$ restricted to $\overline U$.

  3. Define $f(\pi) = \sum_{i\in \pi} f_i$ to be the sum of the frequencies of items in sequence $\pi$.

  4. For $j=1,2,\ldots,n$:
  \hfill {\rm\em (merge)~}

  5. ~~~~With probability ${f(\pi)}/\big({f(\pi)+f(\overline\pi)}\big)$,
  let $i_j \leftarrow \mbox{pop}(\pi)$;
  
  6.~~~~~~otherwise
  let $i_j \leftarrow \mbox{pop}(\overline\pi)$.
  \hfill {\rm\em ({\rm\sf pop$(\pi)$} deletes \& returns the first item remaining in $\pi$)~}
  }}
\caption{A procedure for sampling without replacement, per Lemma~\ref{lemma:merging} (merging).}\label{fig:merge}
  \vspace*{-1ex}
\end{figure*}

We have established that \fcfs is optimally competitive.
In this section we bound its competitive ratio from above.
(Section~\ref{sec:greedy lower bounds} has matching lower bounds.)

Recall that $c_1 \le c_2 \le \cdots \le c_n$.
The cost vector is {\em concave}
if $c_{i+2} - c_{i+1} \le c_{i+1} - c_{i}$ for all $i\in [n-2]$.
Recall $H_j = 1+\frac{1}{2} + \cdots + \frac{1}{j} \sim \ln j$.

\begin{theorem}\label{thm:greedy upper bounds}
  The competitive ratio of the First-Come-First-Served algorithm for Online Slot Allocation is bounded as follows:
  \begin{itemize}
    \item[(i)]
      For general costs, the ratio is at most $1+H_K$,
      where there are $K$ non-maximum coefficients in the cost vector:
      $K=|\{i \in [n] ~|~ c_i < \max_j c_j\}| \le n-1$.
      \item[(ii)]For concave costs, the ratio is at most 2.
        \item[(iii)] 
      For costs $c$ with $c_j = \log_2 j + O(\log \log j)$,
      \fcfs returns an allocation with cost at most $\entropy(f) + O(\log \entropy(f))$,
      where $\entropy(f)$, the entropy of $f/\sum_i f_i$, 
      is a lower bound on $\opt + O(\log \opt)$.
    \end{itemize}
  \end{theorem}
\begin{Proof}
Assume throughout that, for any instance $(f,c)$ of OSA,
the frequency distribution is non-increasing:
$f_1 \ge f_2 \ge \cdots f_n > 0$
(this is without loss of generality, as \fcfs is unbiased).
Hence, $\opt(f,c) = \sum_{j=1}^n c_j f_j$.

\smallskip
\paragraph{Preliminaries.}
Throughout we model \fcfs on $(f,c)$ via the following equivalent random process 
(the compact game from Section~\ref{sec:greedy optimal}):
{\em Generate a random permutation $i_1,i_2,\ldots,i_n$ of $[n]$ by sampling from $f$ without replacement.
For $j=1,2,\ldots,n$, put item $i_j$ in slot $j$.}
The cost of \fcfs on $(f,c)$ is then $\sum_{j=1}^n c_j f({i_j})$.

\smallskip
To bound the probabilities of various events related to this process,
we observe that it can be viewed as a recurse-then-merge process,
similar to mergesort, as shown in Fig.~\ref{fig:merge}.

\begin{lemma}[merging]\label{lemma:merging}
  Let $f$ be a frequency distribution on $[n]$, and let $(U,\overline U)$ be a partition of $[n]$ into two sets.
  Given $f$ and $(U,\overline U)$,
  the random permutation $i_1,\ldots,i_n$ of $[n]$
  generated by the procedure in Fig.~\ref{fig:merge} 
  is distributed as if it were obtained by sampling without replacement from $f$.
\end{lemma}
{
\begin{Proof}
  The case that $U$ or $\overline U$ is empty is trivial,
  so assume otherwise.

  The probability that a given item $i\in U$ is taken first
  is $[f(\pi)/(f(\pi)+f(\overline\pi))] \times [f_i / f(\pi)] = f_i/[f(\pi)+f(\overline\pi)]$.

  The probability that a given item $\overline i \in \overline U$ is taken first
  is $[f(\overline\pi)/(f(\pi)+ f(\overline\pi))] \times [f_{\overline i} / f(\overline\pi)] = f_{\overline i}/[f(\pi)+ f(\overline\pi)]$.

  Thus, the first item is distributed correctly.
  By induction on $n$, the remaining items are distributed
  as a random permutation drawn without replacement, from $f$
  with the first item deleted.
  This gives the correct distribution on the entire permutation.
\end{Proof}
}

It suffices to consider cost vectors in a spanning set:
\begin{lemma}\label{lemma:span}
  If the competitive ratio of \fcfs 
  is at most $\lambda$ for all inputs having cost vectors in some set $X$,
  where each vector in $X$ is non-decreasing,
  then the competitive ratio of \fcfs is also at most $\lambda$ 
  for all inputs having cost vectors in the positive linear span of $X$.
\end{lemma}
{\begin{Proof}
Fix any instance $(f,c')$ where $c'$ is in the positive linear span of $X$,
that is, $c'=\sum_{c\in X} \alpha_c\, c$ where each $\alpha_c\in\Rp$.
By assumption each vector $c\in X$ is non-decreasing, so $c'$ is also.

Let $i_1,i_2,\ldots,i_n$ be a random permutation obtained by sampling without
replacement from $f$.
The expected cost of \fcfs on $(f,c')$ is
(by linearity of expectation)
\[E\Big[\sum_{j=1}^n c'_j \,f_{i_j}\Big]
~=~
\sum_{c\in X} \alpha_c ~E\Big[\sum_{j=1}^n c_j \,f_{i_j}\Big].
\]
Each term $E\big[\sum_{j=1}^n c_j \,f_{i_j}\big]$ 
is the expected cost of $\fcfs$ on instance $(f,c)$,
which by assumption is at most $\lambda\, \opt (f,c) = \lambda\sum_{j=1}^n c_j f_j$.
Thus, the right-hand side above is at most
\medskip

\noindent\hfill $\displaystyle
\sum_{c\in X} \alpha_c ~ \Big[\lambda \sum_{j=1}^n c_j \, f_j\Big]
~=~
\lambda \sum_{j=1}^n c'_j \,f_j
~=~
\lambda\, \opt(f,c')
$ . \hfill
\end{Proof}

}

Now we prove each part in turn.

\subsection*{Part (i) -- general costs.} 
Fix any instance $(f,c)$.  
Recall that $f$ is non-decreasing and $c$ is non-increasing.
Assume further that, for some $k\le K$,
the cost vector $c$ satisfies
$c_1=c_2=\cdots =c_k=0$ 
and $c_{k+1}=\cdots =c_n=1$.
(This is without loss of generality
by Lemma~\ref{lemma:span},
as such cost vectors span all non-decreasing cost vectors 
in $\Rp^n$ with at most $K$ non-maximal coefficients.)
Call items $1,\ldots,k$ {\em large} and the remaining items $k+1,\ldots,n$ {\em small}.
Let $\epsilon$ denote \opt's cost, $\epsilon=\sum_{i=k+1}^n f_i$,
the sum of the small items' frequencies.

Let $\calI=(i_1,\ldots,i_n)$ of be a random permutation of the items
obtained by sampling from $f$ without replacement.
For the analysis, generate the random permutation $\calI$ via the process described in Lemma~\ref{lemma:merging}:
choose a random permutation
$\calL = (\ell_1,\ldots,\ell_k)$ 
of the large items 
(sampling from $f$ restricted to large items);
choose a random permutation $\calS$ of the small items 
(sampling from $f$ restricted to small items);
then, in the {\em merge phase},
for each $j=1,2,\ldots,n$,
obtain $i_j$
by taking either the first remaining large item
(with probability proportional to the frequency of the remaining large items)
or the first remaining small item
(with probability proportional to the frequency of the remaining small items).

The small items can't contribute more than $\epsilon$ to the cost of \fcfs.
We focus just on the large items, and show that they contribute in expectation at most $\epsilon H_k$.

\Fcfs pays for any item that isn't chosen within the first $k$ iterations of the merge phase.
Focus on just these iterations:
fix any $j\le k$ and consider the start of iteration $j$.
Let $h$ be the number of large items chosen so far;
these items will definitely be free. 
Further, \fcfs will definitely pay for $p$ large items,
where $p=j-1-h$ is 
the number of not-yet-chosen large items
(i.e., $k-h$)
minus the number of iterations left within the first $k$
(i.e., $k-j+1$).
The $h$ free items are first in $\calL$; the $p$ to-be-paid items are last:
\newcommand{\YES}[1]{\shortstack{\scriptsize 0\\$#1$}}
\newcommand{\NO}[1]{\shortstack[l]{\text{\scriptsize \$}\\$#1$}}
\newcommand{\MB}[1]{\shortstack[l]{{\scriptsize ?}\\$#1$}}
\renewcommand{\MB}[1]{?}
\renewcommand{\YES}[1]{#1}
\renewcommand{\NO}[1]{#1}
\begin{equation}\label{eqn:Lj}
\calL = (
\underbrace{\YES{\ell_1}, \YES{\ell_2}, \ldots, \YES{\ell_{h}}}_{h \text{ free}}, 
\,
\underbrace{\MB{\ell_{h+1}},\,\MB{\ell_{h+2}},\, \ldots\,,\, \MB{\ell_{k-p}}}_{\text{not determined}\!}, \,
\underbrace{\NO{\ell_{k-p+1}}, \ldots, \NO{\ell_k}}_{p \text{ paid}}).
\end{equation}
During iteration $j$, the status (paid or free) of exactly one large item will be determined.
Define $X_j$ to be the corresponding contribution to the cost:
If iteration $j$ chooses a large item, $\ell_{h+1}$, 
then that item becomes {\em free}, and $X_j$ is zero.
Otherwise, the last not-determined item, $\ell_{k-p}$, becomes {\em paid},
and $X_j$ is that item's frequency, $f(\ell_{k-p})$.

\Fcfs pays in total $\sum_{j=1}^k X_j$ for the large items.

Let $\Phi_j(F,P)$ denote the event that, at the start of iteration $j$
of the merge phase, the sequence of free items is $F$
and the sequence of paid items is $P$.
Let $N$ denote the unordered set of not-determined large items
(abusing notation, $N = [k]-F-P$).
Define $f(N) = \sum_{i\in N} f_i$.

\begin{lemma}\label{lemma:smallprob}
  For any $(F,P)$,
  the probability that item $i_j$ is small,
  conditioned on $\Phi_j(F,P)$,
  is at most $\epsilon/f(N)$.
\end{lemma}
\begin{Proof}
  Condition further on $\calS = S$ and $\calL = L$,
  where $S$ is any permutation of the small items
  and $L$ is any permutation of the large items that starts with $F$ and ends with $P$.
  With $\Phi_j(F,P)$, this completely determines
  the entire state at the start of iteration $j$ of the merge phase,
  including the outcomes of iterations $1,\ldots,j-1$.
  Starting in any such state, the probability that iteration $j$ chooses a small item
  is the following ratio:
  the total frequency of the not-yet-chosen small items 
  (the last $n-k-|P|$ items in $S$),
  divided by the total frequency of all not-yet-chosen items
  (the last $n-k-|P|$ in $S$ and the last $k-|F|$ in $L$).
  The numerator is at most $\epsilon$;
  the denominator is at least $f(N) + f(P) \ge f(N)$.
  So, for any $(S,L)$,
  \\\centerline{$\Pr[i_j \text{ small}~|~\calS = S;\, \calL = L;\, \Phi_j(F,P)] ~\le~ \epsilon/f(N)$.}

  So $\Pr[i_j \text{ small}~|~\Phi_j(F,P)] \le \epsilon/f(N)$.
\end{Proof}

\begin{lemma}\label{lemma:condexp}
  For any $(F,P)$,
  the conditional expectation of $X_j$,
  given $\Phi_j(F,P)$ and that item $i_j$ is small,
  is at most $f(N)/|N|$.
\end{lemma}
\begin{Proof}
  Let $\Psi_j$ denote the event that $\Phi_j(F,P)$ occurs and $i_j$ is small.
  This event determines that the large-item permutation $\calL$ 
  has $F$ as a prefix and $P$ as a suffix.

  We claim that this is the only way that it conditions $\calL$:
  that is, the conditional distribution of $\calL$ 
  is the same as that of a random permutation of the large items
  that is obtained just by sampling with repetition from $f$,
  conditioned on having $F$ as a prefix and $P$ as a suffix.
  Here's why:
  Fix any ordering $S$ of the small items,
  and any two orderings $L$ and $L'$ of the large items
  that are consistent with $\Psi_j$  ($L$ and $L'$ order $N$ differently).
  All choices made in iterations $1,\ldots,j$ of the merge phase
  are independent of $N$'s order,
  so 
  \\[3pt]\centerline{$\Pr[\Psi_j~|~ \calS=S;\,\calL=L]~=~\Pr[\Psi_j~|~\calS=S;\,\calL=L'].$}
  \\[3pt]This holds for any $S$, so
  \\[3pt]\centerline{$\Pr[\Psi_j~|~ \calL=L] ~=~ \Pr[\Psi_j~|~\calL=L'].$}
  \\[3pt]
  That is, $\Phi_j$ reveals no information about the ordering of $N$ within $\calL$.
  This proves the claim.\footnote
  {Here's a sketch of a more detailed proof of this conclusion.
    Let $\calU$ be the set of large-item permutations that start with $F$ and end with $P$.
    For $L \not\in \calU$, 
    $\Pr[\calL = L~|~\Psi_j] = \Pr[\calL = L ~|~\calL \in \calU] = 0$.
    For $L,L' \in \calU$
    the equation before the footnote implies (by calculation)
     \[
    \frac{\Pr[\calL = L~~|~\Psi_j]}{\Pr[\calL = L'~|~\Psi_j]}
    =
    \frac{\Pr[\calL = L]}{\Pr[\calL = L']}
    =
    \frac{\Pr[\calL = L~~|~\calL\in \calU]}{\Pr[\calL = L'~|~\calL\in \calU]}.
    \]
    Hence, for every $L$,
    $\Pr[\calL = L~|~\Psi_j] = \Pr[\calL = L~|~\calL\in \calU]$.
  }

  Now recall that $\ell_{|F|+1}, \ldots, \ell_{k-|P|}$ denotes the order of $N$ within $\calL$,
  so that $X_j= f(\ell_{k-|P|})$.
  By Observation~\ref{obs:decreasing expectations}, 
  even conditioning $\calL$ on $F$ and $P$,
  we have 
  $E[f(\ell_{|F|+1})] \ge E[f(\ell_{|F|+2})] \ge \cdots \ge E[f(\ell_{k-|P|})]$.
  The sum of these $|N|$ expectations is $f(N)$,
  so the last (and minimum) one is at most their average $f(N)/|N|$.
\end{Proof}
The two lemmas imply that, for any $(F,P)$, the conditional expectation $E[X_j\,|\,\Phi_j(F,P)]$
is at most the product $\big(\epsilon/f(N)\big)\big(f(N)/|N|\big) = \epsilon/|N|$.
Since $|N|$ is necessarily $k-j+1$, this implies
$E[X_j] \le \epsilon/(k-j+1)$.
Thus, the large items cost in expectation
at most $\sum_{j=1}^k E[X_j] \le \sum_{j=1}^k \epsilon/(k-h+1) = \epsilon H_k$.

\smallskip
This proves Part (i) of Thm.~\ref{thm:greedy upper bounds}.

\subsection*{Part (ii) -- concave costs.}
Fix any instance $(f,c)$ where $c$ is concave and $f$ is non-decreasing.
Assume $c_1 = 0$
(without loss of generality, otherwise, subtracting $c_1$ from all costs decreases the cost
of any allocation by $n\, c_1$, only improving the competitive ratio).
Assume that, for some integer $k\ge 2$,
each $c_{j} = \min(j, k)-1$
(without loss of generality, by Lemma~\ref{lemma:span},
as such vectors span all concave cost vectors with $c_1=0$).

Generate random permutation $i_1,\ldots,i_n$ by drawing without replacement from $f$.
Let $j_i$ denote the position of item $i$ in the permutation
(and the slot \fcfs puts $i$ in, at cost $c_{j_i}=\min(j_i,k)-1$).
The cost of \fcfs is upper-bounded by
\(
\sum_{i=1}^{k-1} (j_i-1)f_i
+\sum_{i=k}^{n} (k-1)f_i.
\)

For any two items $h,i\in[n]$,
applying Lemma~\ref{lemma:merging} (merging)
to the two singleton lists $(h)$ and $(i)$,
the probability that $h$ occurs before $i$ in the random permutation is $f_{h}/(f_i+f_{h})$.
Hence, for $i \in [k-1]$, the term $(j_i-1)$ in the upper bound,
which equals the number of items drawn before item $i$, has expectation
\[
{\sum_{{h} \ne i} \Pr[j_{h} < j_i]}
 ~=~
\sum_{{h} \ne i} \frac{f_{h}}{f_i + f_{h}}
~\le~
i-1 + \sum_{h=i+1}^{n} \frac{f_h}{f_i}.
\]
Hence, the expectation of the upper bound is at most
\[
\sum_{i=1}^{k-1} \Big(i-1~~+\sum_{h=i+1}^{n} \frac{f_h}{f_i}\Big)f_i
~~+\sum_{i=k}^n (k-1)f_i.
\]
Canceling $f_i$'s and simplifying, this is
\begin{eqnarray*}
  \lefteqn{\sum_{i=1}^{k-1} (i-1)  f_i ~+\sum_{h=2}^{n} f_h \sum_{i=1}^{\!\!\min(h,k-1)\!\!} 1 ~+ \sum_{i=k}^n (k-1)f_i}
\\&=& 2\sum_{h=1}^{k-1} (h-1)  f_i~+2\sum_{h=k}^n (k-1)f_h
~~=~2\,\opt(f,c).
\end{eqnarray*}

\subsection*{Part (iii) -- logarithmic cost.}
Fix any instance $(f,c)$ with $c_j = \log_2 j + O(\log\log j)$.
Assume to ease notation that $\sum_i f_i = 1$. 
Let $i_1,\ldots,i_n$ be a random permutation obtained by drawing from $f$ without replacement.  
Let $j_i$ be the location of $i$ in the permutation.

\begin{lemma}\label{lemma:log}
For any item $i$ and concave function $F$, $E[F(j_i)] \le F(1/f_i)$.
\end{lemma}
\begin{Proof}
  For any other item $h\ne i$,
  applying Lemma~\ref{lemma:merging} (merging)
  to the two singleton lists $(h)$ and $(i)$,
  the probability that $h$ occurs before $i$ in the random permutation is $f_{h}/(f_i+f_{h})$.
  Since $j_i-1$ is the number of items drawn before item $i$,
  by linearity of expectation,
  \[
  E[j_i] ~=~ 1+\sum_{j\ne i} \frac{f_j}{f_i+f_j} ~\le~ 1 + \frac{1-f_i}{f_i} ~=~ \frac{1}{f_i}.
  \]
  $F$'s concavity gives $E[F(j_i)] \le F(E[j_i]) \le F(1/f_i)$.
\end{Proof}

Recall $c_j = \log_2 j + O(\log\log j)$.
By Lemma~\ref{lemma:log} and the concavity of the logarithm,
the expected cost of \fcfs is
\begin{eqnarray*}
\sum_{i=1}^n  f_i\, E[c({j_i})]
&=&\sum_{i=1}^n f_i\, E[\log_2 j_i + O(\log\log j_i)]
\\&\le&
\sum_{i=1}^n f_i\, \Big[\log_2 \frac{1}{f_i} + O\Big(\log\log \frac{1}{f_i}\Big)\Big]
\\&=&
\entropy(f) + O(\log \entropy(f)).
\end{eqnarray*}
(Above $\entropy(f)$ is the entropy of $f$.)

The minimum cost is $\opt = \sum_{i=1}^n c_i f_i$.
By Kraft's inequality~\cite{kraft1949device} and a calculation, there is a prefix-free code
of cost $\sum_{i=1}^n f_i (c_i + O(\log c_i)) = \opt + O(\log \opt)$.
Since that code is prefix-free, its cost is at least the entropy of $f$,
so $\entropy(f) \le \opt + O(\log \opt)$.
\end{Proof}

\section{Lower bounds on optimal competitive ratio}\label{sec:greedy lower bound}\label{sec:greedy lower bounds}
In this section we prove the following tight lower bounds
on the optimal competitive ratios of any online algorithm for OSA:

\begin{theorem}\label{thm:greedy lower bounds}~For Online Slot Allocation:
  \begin{itemize}
    \item[(i)]
      For 0/1 costs with at most $K$ non-zero coefficients
    ($\{|i\in[n]~|~c_i \ne \max_j c_j\} \le K$),
    the optimal competitive ratio is at least $1+H_K$ (where $H_{K} = 1 + \frac{1}{2} + \frac{1}{3} + \cdots +\frac{1}{K} \sim \ln K$).
    \item[(ii)] For concave 0/1 costs, the optimal competitive ratio is at least 2.
    \\
  \end{itemize}
\end{theorem}
\begin{Proof}
  We prove bounds (i) and (ii) for \fcfs.
  Since \fcfs is optimally competitive, the theorem follows.

\subsection*{Part (i) -- general costs.}
Fix arbitrarily large $n$ and let $\epsilon\rightarrow 0$.
Consider the instance $(f,c)$ 
whose cost vector $c\in\{0,1\}^n$ has $K$ zeros,
where the first $K$ frequencies equal 1 (call them  {\em large})
and the last $n' = n-K$ frequencies equal $\epsilon/n'$ (call them {\em small}).
The expected cost that \fcfs pays for the large items is at least 
the probability that some large item ends up in position $K+1$ or higher,
which equals the probability that some small item comes before some large item.
Applying Lemma~\ref{lemma:merging} (merging)
to merge the small item into the large ones,
the probability that all small items come {\em after} all $K$ large items is
\begin{eqnarray*}
&&\prod_{\ell=0}^{K-1} \frac{K-\ell}{K-\ell + \epsilon}
~=~
\prod_{\ell=0}^{K-1} ~1 - \frac{\epsilon}{K-\ell + \epsilon}
\\ &&=
1 - \sum_{\ell=0}^{K-1} \frac{\epsilon}{K-\ell + \epsilon} + O(\epsilon^2)
~=~
1 - \epsilon H_{K} + O(\epsilon^2).
\end{eqnarray*}
Hence, the probability that some small item comes before some large item
(a lower bound on \fcfs's expected cost for the large items)
is $\epsilon H_{K} - O(\epsilon^2)$.

\Fcfs's cost for the small items is at least $(n'-K) \epsilon/n' = \epsilon -O(\epsilon K /n)$.
Thus, \fcfs total expected cost is at least $\epsilon [ 1 + H_K - O(K/n + \epsilon)]$.

The minimum allocation cost is $\epsilon$.
The ratio tends to $1+H_{K}$ (as $\epsilon\rightarrow 0$ and $n/K\rightarrow\infty$),
proving Part (i).

\subsection*{Part (ii) -- concave costs.}
Fix any $n$.
Consider the instance $(f,c)$ with
cost vector $c=(0,1,1,\ldots,1)$ 
and frequency vector $f=(1,\epsilon,\epsilon, \ldots,\epsilon)$,
where $\epsilon\rightarrow 0$.
Let $i_1,\ldots,i_n$ be a random permutation obtained by drawing without replacement from $f$.  \Fcfs's expected cost is {\em at least}
\begin{eqnarray*}
\lefteqn{\Pr[i_1 \ne 1]\cdot 1 \,+\, \Pr[i_1 = 1]\cdot (n-1)\epsilon}
\\ &=& \frac{(n-1)\epsilon}{1+(n-1)\epsilon} + \frac{(n-1)\epsilon}{1+(n-1)\epsilon}
 ~=~ \frac{2 (n-1)\epsilon}{1+(n-1)\epsilon}.
\end{eqnarray*}
The minimum allocation cost is $(n-1)\epsilon$.
The ratio is $2/(1+(n-1)\epsilon)$,
which tends to 2 as $\epsilon\rightarrow 0$.
\end{Proof}

\section{Online Huffman coding (OHC)
}\label{sec:huffman}

In this section we prove the 
following performance guarantee
for a \fcfs algorithm for OHC:

\begin{theorem}\label{thm:greedy huffman}
  There exists an algorithm for Online Huffman Coding such that,
  for any instance $f$, the algorithm returns a prefix-free code
  of expected cost 
  at most $\entropy(f) + 2\log_2(1+ \entropy(f)) + 2 \le 5\,\entropy(f)$,
  where $\entropy(f)$, the entropy of $f/\sum_{i=1}^n f_i$, is a lower bound on $\opt$.
\end{theorem}

Online Huffman Coding is not a special case of Online Slot Allocation because of the prefix-free constraint:
assigning a codeword $j$ to a given item 
precludes the use of other codewords (those with $j$ as a prefix).
To work around this, the algorithm uses a so-called {\em universal (codeword) set}
--- an infinite prefix-free subset $\calU$ of $\{0,1\}^*$ ---
to effectively relax the prefix-free constraint.

We first fix a particular universal codeword set with ``small'' codewords.
Let $c_\calU(j)$ be the length of the $j$th smallest string in $\calU$.
Call $c_\calU$ the {\em cost function} for $\calU$.

\begin{lemma}\label{lemma:universal}
  There is a universal set $\calU$ with cost 
  $c_\calU(j) = \lfloor 2 + \log_2 j + 2 \log_2 (1+\log_2 j) \rfloor$.
\end{lemma}
\begin{Proof}
  By calculation, the cost function satisfies
  $\sum_{j=1}^\infty 1/2^{c_\calU(j)}\le 1$.
  The existence of the prefix-free set
  $\calU$ follows by Kraft's inequality~\cite{kraft1949device}.
\end{Proof}
(For concreteness, here is a more explicit description of $\calU$,
following e.g.,~\cite{golin2002huffman,golin2012huffman}.
For each string $x\in\{0,1\}^+$,
add the string $wx$ to $\calU$,
where $w$ is as computed as follows.
Let $\ell$ be the binary encoding of the length of $x$.
Delete the first ``1'', replace each ``0'' or ``1''
by ``00'' or ``11'', respectively, then append a ``01''.
This gives $w$.
$\calU$ is prefix-free because $w$ encodes the length $\ell$ of $x$
in a prefix-free fashion, and $\ell$ determines when $x$ ends.)

For the rest of the section, fix $\calU$ and $c_\calU$ from Lemma~\ref{lemma:universal}.
Define the Online Huffman Coding algorithm \fcfsu as follows:
{\em Given a new item $i$, 
  allocate the next smallest unused codeword in $\calU$ to item $i$.}
\fcfsu returns a prefix-free code because $\calU$ is prefix-free.

Note that \fcfsu is \fcfs applied to the OSA instance $(f,c_\calU)$,
which is equivalent to the OHC instance $f$ with the additional
constraint that codewords must be chosen from $\calU$.
The $j$th smallest string in $\{0,1\}^+$
has length $c_\calU(j) - O(\log\log j)$,
so it is easy to show that the added constraint
increases \opt by at most an additive $O(\log \opt)$.
This observation
and Thm.~\ref{thm:greedy upper bounds}, Part (iii),
imply the following looser performance guarantee:
\begin{lemma}
  For any instance $f$ of Online Huffman Coding,
  the prefix-free code returned by
  \fcfsu has expected cost at most $\opt(f) + O(\log \opt(f))$.
\end{lemma}

But a tighter, more direct analysis proves Thm.~\ref{thm:greedy huffman}:
\begin{Proof}(Thm.~\ref{thm:greedy huffman}.)
  Assume without loss of generality that $\sum_i f_i = 1$.
  Let $i_1,\ldots,i_n$ be a random permutation 
  generated by drawing without replacement from $f$.  
  Let r..~$j_i$ be the position of $i$ in the permutation.

  By Lemma~\ref{lemma:log}
  and the concavity of the logarithm, $E[\log_2 j_i] \le \log_2(1/f_i)$.
  Summing over $i$ gives
  \[\sum_{i=1}^n f_i\, E[\log_2 j_i] ~\le~ \sum_{i=1}^n f_i \log_2 \frac{1}{f_i} ~=~ \entropy(f).\]

  Using this inequality (twice),
  the choice of $c_\calU$ in Lemma~\ref{lemma:universal},
  and the concavity of log (twice),
  the expected cost of the allocation, $\sum_{i=1}^n f_i \,c_\calU(j_i)$, is
  \[
    2 + \Big(\sum_{i=1}^n f_i\, E[\log_2 j_i]\Big)
  +~ 2\log_2\big(1+\textstyle \sum_{i=1}^n f_i\, E[\log_2 j_i]\big)
  \]
  \nopagebreak
  $~\le~
  2 + \entropy(f) + 2\log_2(1+\entropy(f)).$
\end{Proof}

\bibliographystyle{abbrv}
\bibliography{bib}

\end{document}